# Giant optical nonlinearity in single silicon nanostructure: ultrasmall all-optical switch and super-resolution imaging

*Yi-Shiou Duh[1], *Yusuke Nagasaki[2], Yu-Lung Tang[1], Pang-Han Wu[1], Hao-Yu Cheng[3], Te-Hsin Yen[1], Hou-Xian Ding[1], Kentaro Nishida[2,4], Ikuto Hotta[2], Jhen-Hong Yang[5], Yu-Ping Lo[6], Kuo-Ping Chen[6], Katsumasa Fujita[2,4], Chih-Wei Chang[7], Kung-Hsuan Lin[3], Junichi Takahara[2,8], and Shi-Wei Chu[1,9]
*These authors contributed equally to this work

[1]Department of Physics, National Taiwan University, 1, Sec 4, Roosevelt Rd., Taipei 10617, Taiwan
[2]Graduate School of Engineering, Osaka University, 2-1 Yamadaoka, Suita, Osaka 565-0871, Japan
[3]Institute of Physics, Academia Sinica, 128, Sec. 2, Academia Rd., Taipei 11529, Taiwan
[4]AIST-Osaka University Advanced Photonics and Biosensing Open Innovation Laboratory, AIST, 2-1 Yamadaoka, Suita, Osaka 565-0851, Japan
[5]Institute of Photonic System, National Chiao Tung University, 301 Gaofa 3rd Road, Tainan 711, Taiwan
[6]Institute of Imaging and Biomedical Photonics, National Chiao Tung University, 301 Gaofa 3rd Road, Tainan 711, Taiwan
[7]Center for Condensed Matter Sciences, National Taiwan University, 1, Sec 4, Roosevelt Rd., Taipei 10617, Taiwan
[8]Photonics Center, Graduate School of Engineering, Osaka University, 2-1 Yamadaoka, Suita, Osaka 565-0871, Japan
[9]Molecular Imaging Center, National Taiwan University, 1, Sec 4, Roosevelt Rd., Taipei 10617, Taiwan
Corresponding author: Kung-Hsuan Lin (linkh@sinica.edu.tw); Junichi Takahara (takahara@ap.eng.osaka-u.ac.jp); Shi-Wei Chu (swchu@phys.ntu.edu.tw)

## Abstract

Silicon photonics has attracted significant interest in recent years due to its potential in integrated photonics components[1,2] as well as all-dielectric meta-optics elements.[3] Strong photon-photon interactions, aka optical nonlinearity, realizes active control of aforementioned photonic devices.[4,5] However, intrinsic nonlinearity of Si is too weak to envision practical applications. To boost the nonlinear response, long interaction-length structures such as waveguides, or resonant structures such as microring resonators or photonic crystals have been adopted.[6,7] Nevertheless, their feature sizes are typically larger than 10 µm, much larger than their electronic counterparts. Here we discover, when reducing the size of Si resonator down to ~100 nm, a giant photothermal nonlinearity that yields 400% reversible and repeatable deviation from linear scattering response at low excitation intensity (mW/µm$^2$). The equivalent nonlinear index $n_2$ at nanoscale is five-order larger than that of bulk, due to Mie resonance enhanced absorption and high-efficiency heating in the thermally isolated nanostructure. In addition, the nanoscale thermal relaxation time reaches nanosecond, implying GHz modulation speed. This large and fast nonlinearity enables applications toward all-optical control in nanoscale, as well as super-resolution imaging of silicon.

## Main

Silicon, because of its natural abundance and its compatibility with industrial production lines, is the most widely used material in modern electronics industry. Yet, due to its indirect bandgap, Si has limited applications in photonics. It is a long awaited goal to amalgamate photonics with the advantages of silicon. It was not until the recent decade had we witnessed dramatic progresses in silicon photonics, including amplification, lasing, supercontinuum generation, etc.[1,2,5]

One major challenge for active silicon photonics is optical nonlinearity that can implement desirable electron-photon or photon-photon interactions within a device. Conventionally, Kerr-type nonlinearities provide ultrafast response. but the magnitude of nonlinearity ($n_2$) is on the order of $10^{-9}$ µm$^2$/mW. Photothermal effects are known to provide much larger nonlinear response, with effective nonlinear coefficient $n_2$ approaches $10^{-6}$ µm$^2$/mW.[8] From a simple estimation based on $n = n_0 + n_2 I$, in which n is the overall refractive index, $n_0$ is the linear index, $n_2$ is the nonlinear index, and I is excitation intensity, to create 10% nonlinear deviation from the linear response, excitation intensity on the order of $10^5$ mW/µm$^2$ is required.

Because of the weak nonlinearity of silicon, typical design of nonlinear silicon photonic components requires resonant structures such as microring resonators and photonic crystals.[6,7] To realize high-Q resonance, the feature size of these resonant structures are on the order of 10 µm. For example, on-chip single-layer integration of silicon electronics and photonics elements was demonstrated very recently,[9] but the photonics component, which was a microring resonator, was much larger than the electronic transistors.

Inspired by metal-based plasmonics, an emerging field is to significantly enhance light-matter interactions via strong light confinements utilizing nanoscale high-index dielectrics,[10] without being compromised by metal loss. The meta-silicon-material has led to various unexpected optical properties, e.g. Mie-resonance-induced localization as well as electric/magnetic dipole/multipoles,[4] optical magnetism,[11] directional emission,[12] broadband perfect reflector,[13] high-efficiency hologram,[14] optical topological states,[15] multicolor nano-display,[16] etc. Nanostructured Si has also displayed many unusual optical control or nonlinearity.[17] For example, silicon metasurfaces have led to ultracompact phase controller[18] and five-order enhancement of third harmonic generation,[19] but their feature sizes are still much larger than wavelength. In a single silicon nanoparticle, 100-fold enhancement of third harmonic generation[20] and two-photon absorption[21] were observed recently. However, the required intensity is ~ $10^4$ mW/$\mu m^2$, and the corresponding nonlinear deviation is only less than 1%, i.e. the optical nonlinearities are still far from sufficient to realize applications such as all-optical control at low light intensity.

In this work, we combine Mie resonance with photothermal effect, and report unexpectedly large photothermal nonlinearity in a single silicon nanostructure of ~0.001 $\mu m^3$ volume. The single-nanostructure nonlinearity enables 400% enhancement or 70% reduction of scattering, i.e. significant deviation from linear response, at excitation intensity of merely 1 - 10 mW/$\mu m^2$. The equivalent $n_2$ reaches $10^{-1}$ $\mu m^2$/mW, which is five orders larger than photothermal nonlinearity of bulk silicon. Furthermore, thermal relaxation time of the nanostructure-based nonlinear response is on the order of nanosecond, leading to the potential of GHz operation. The large nonlinearity and the fast response hold great potential for all-optical nano-silicon applications.

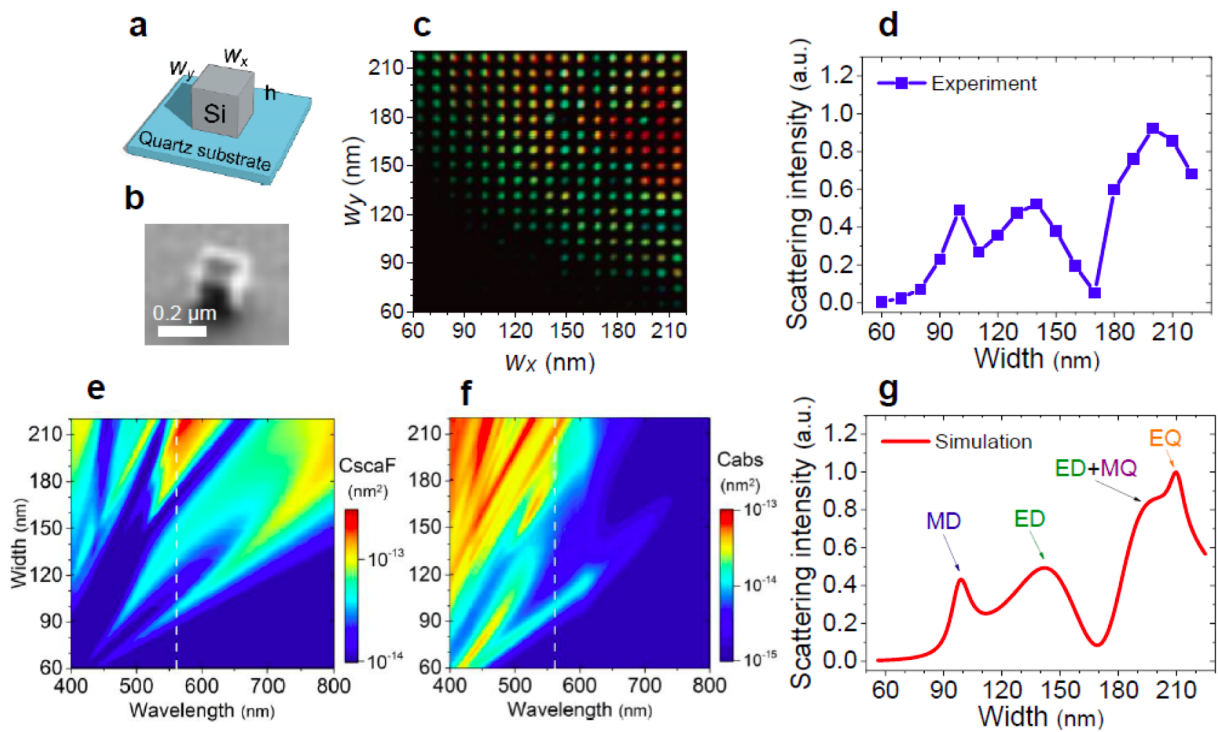

**Fig. 1 | Scattering property of Si nanostructures.**
**a.** We use Si nanoblocks on a quartz substrate, whose lateral dimensions ($w_x$ and $w_y$) vary from 60 nm to 220 nm, with a 10-nm step, and a fixed height (h) of 150 nm. **b.** Focused ion beam image of a typical Si nanoblock with $w_x = w_y = 200$ nm. Resolution is compromised due to the charging effect of the non-conductive quartz substrate. **c.** Halogen lamp illuminated dark-field image of isolated Si nanoblocks, showing colorful resonance among different sizes. **d.** 561-nm scattering intensity along the array diagonal ($w = w_x = w_y$), presenting the characteristic multipole resonance of silicon nanostructures. The excitation intensity is 1.3 mW/µm$^2$. **e.** Simulated single-particle forward scattering and **f.** absorption cross section spectra versus different nanoblock widths. The white dashed line marks the 561-nm laser, which excites multiple resonances. **g.** Simulation of size-dependent 561-nm scattering, agreeing well with experiment results. MD: magnetic dipole; ED: electric dipole; EQ: electric quadrupole; MQ: magnetic quadrupole.

The sample is a single-crystalline Si nanoblock array on quartz (see Fig. 1a and fabrication detail in Methods), which has been demonstrated recently to host multipolar electric/magnetic resonances within a single unit.[16,22] Fig. 1b is a scanning ion-beam microscope image of one Si nanoblock, showing the high-quality sharp edges and corners. In the array, nanoblock height is fixed at 150 nm, and lateral dimensions ($w_x$ and $w_y$) increase from 60 to 220 nm in 10-nm step, to induce controllable wavelength shift of Mie resonance, as shown in the colorful Fig. 1c (see setup in Methods and Fig. E1). Based on the transparent quartz substrate and a 561-nm dark-field laser scanning microscope, Fig. 1d shows size-dependent scattering intensity along the diagonal nanoblocks ($w_x = w_y$). The choice of this wavelength allows induction of multipole Mie resonances, as shown by the white lines in the size-dependent spectra of Fig. 1e and 1f (see Fig. E2 for multipole decomposition analysis). Fig. 1g is simulated size-dependent scattering at 561 nm, agreeing well with Fig. 1d. Additional theory-experiment correspondences on single nanoblock spectrum is given in Fig. E3. It is interesting to notice that by varying the nanoblock size, more than 80% scattering intensity variation is observed. A more interesting question would be whether similar variations could be found by optically tuning silicon's refractive index, not size, to reach unprecedented optical nonlinearity.

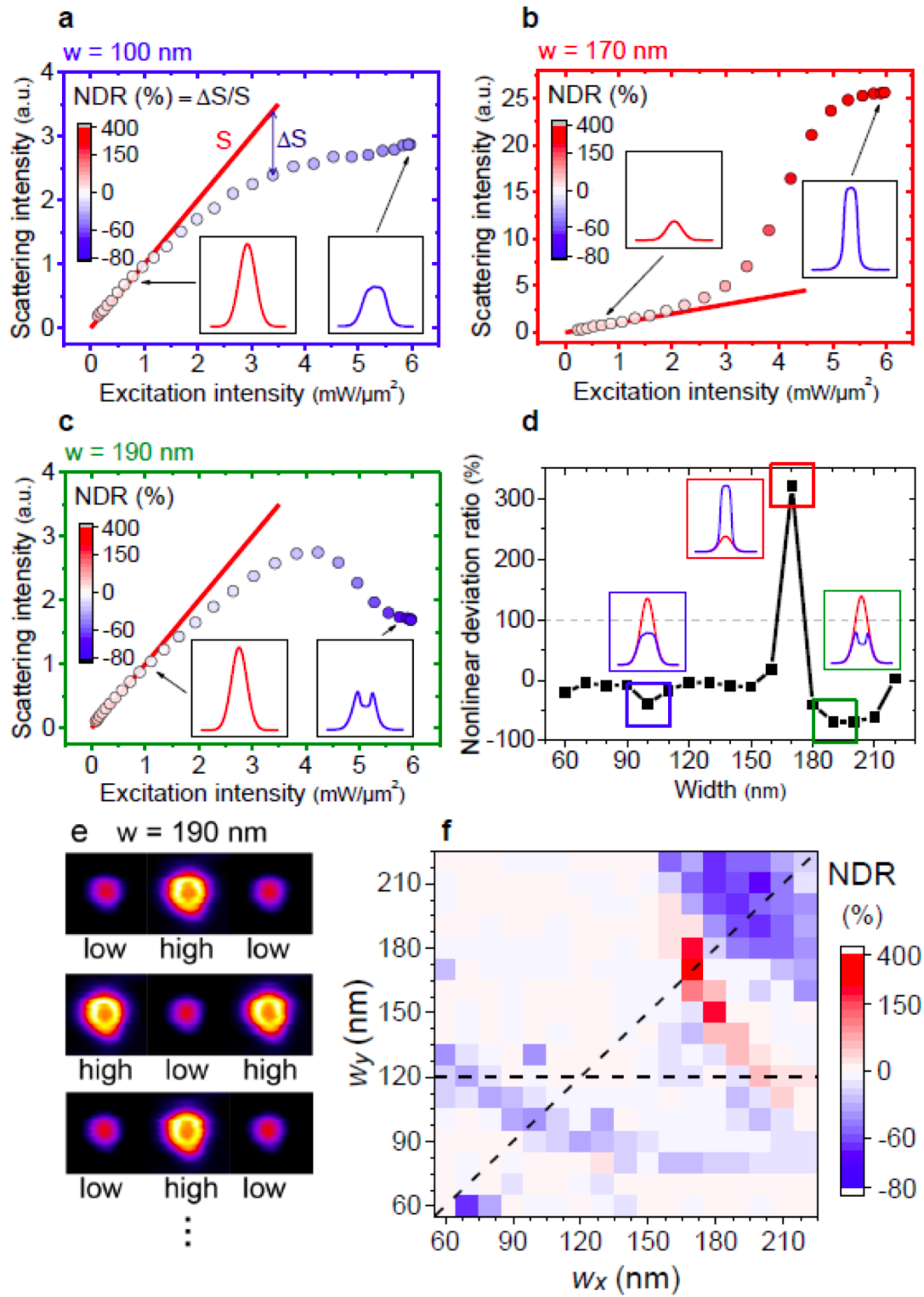

**Fig. 2 | Experimental observation of nonlinear scattering.**

**a, b, c.** Excitation intensity dependent scattering for *w* = 100 nm, 170 nm, 190 nm, respectively, observed by a dark-field laser scanning microscope at *λ* = 561 nm. In the main frames, red lines and colored dots indicate linear scattering intensity (S, extrapolated from low-intensity excitation, see Fig. E4 & E5) and measured scattering intensity (whose deviation from S is ΔS), respectively. Unexpectedly large nonlinearity is manifested through significant deviation from linear intensity dependence. The color of the dots represents nonlinear deviation ratio (NDR), i.e. percentage of ΔS/S. The insets are corresponding PSFs at low and high intensities (lateral distance: 4μm). **d.** NDR versus different size of nanoblocks at 6 mW/μm$^2$ excitation intensity. A dotted line marks 100%, which means no nonlinear response. The blue, red, and green rectangles highlight the regions of large NDR. The insets present corresponding PSF profiles at low excitation (red curves) and high excitation (blue curves) intensities, where large nonlinearity is again manifested by the large deviation of blue profiles from Gaussian distribution. **e.** The PSF recovery during repetitive switching between low-intensity (1.3 mW/μm$^2$) and high-intensity (6.0 mW/μm$^2$) excitations, demonstrating reversible and repeatable nonlinear responses (see Fig. E6 for other nanoblocks). **f.** Experimental NDR map of the whole array. The diagonal dashed line marks the nanoblocks for analysis in Fig. 1d and Fig. 2. The horizontal dashed line marks non-diagonal nanoblocks that are compared to simulation in Fig. E7

The nonlinearity of Si nanoblock scattering is studied via a laser-scanning microscope (xy-scan, see supplementary Methods) at 561-nm, and the results are shown in Fig. 2. The concept of xy-scan to characterize nanostructure nonlinearity is adapted from z-scan. In z-scan, sample should be much thinner than axial (z) length of focus, and when a focused beam scans across the thin sample, deviation from a linear trend indicates the existence of nonlinearity. Here in xy-scan, the size of nanoblocks is much smaller than lateral (xy) point spread function (PSF), and when a focused laser beam scans laterally across a nanoblock, deviation from a Gaussian profile indicates nonlinear response,[23] as shown by the insets of Fig. 2a-2d.

In Fig. 2a, which corresponds to the *w* = 100 nm nanoblock (magnetic dipole dominates, the first peak in Fig. 1d), at low intensity up to 1.5 mW/μm$^2$, the scattering response is linear. Nevertheless, as the laser intensity increases, scattering starts to saturate, i.e. negatively deviates from a linear trend (red line). We define nonlinear deviation ratio (NDR) = ΔS/S, where ΔS is the percentage deviation of measured scattering, and S is the extrapolated linear response. Accordingly, NDR of -50% is obtained at 6 mW/μm$^2$, whose PSF significantly deviates from the original Gaussian profile (see inset of Fig. 2d, which also shows the -50% NDR).

Fig. 2b presents the scattering nonlinearity of the w = 170 nm nanoblock, whose scattering is relatively weak and off-resonance in Fig. 1d. Similar to Fig. 2a, scattering is linear at low excitation intensity, but quite differently at high intensity, the scattering signal of this weak-scattering particle exhibits a sharp reverse saturation, i.e. increase with a very large slope, and then saturates. At 5 mW/μm$^2$, more than 400% positive NDR is observed.

Fig. 2c shows the case of the w = 190 nm nanoblock, which exhibits the largest scattering intensity in Fig. 1d. Again, scattering is linear at low intensity, and at high intensity, the scattering response exhibits a negative deviation, similar to Fig. 2a.

Nevertheless, here scattering is "super-saturated", i.e. reduced with increasing excitation, and the minimum NDR reaches -70%.

Apparently, each particle shows different nonlinear responses, and the NDRs at 6 mW/μm$^2$ are summarized in Fig. 2d, with corresponding PSF profiles. The reversibility and repeatability of the nonlinear responses are demonstrated in Fig. 2e and Fig. E6, thus excluding the possibility of photodamage or oxidation of silicon. The size-dependent NDR of the whole array is given in Fig. 2f, leading to a few insights.

First, these large nonlinear deviations at relatively low excitation intensity lead to an effective nonlinear index $n_2$ of $10^{-1}$ μm$^2$/mW (see methods and Fig. E8 for derivation). This value is much larger than the reported photothermal nonlinearity of silicon ($n_2$ ~ $10^{-6}$ μm$^2$/mW), featuring five-order improvement with an ultrasmall mode volume of 0.001 μm$^3$.

Second, comparing Fig. 2d to Fig. 1d, the positive and negative NDR values in Fig. 2d correspond well to the valley and peaks in Fig. 1d, but not vice versa. For instance, the w = 140 nm Si nanoblock is the second resonance peak in Fig. 1d, but its NDR is less than 5%, much smaller than that of other two peaks. The w = 110 nm Si nanoblock exhibits a scattering valley in Fig. 1d, but no nonlinearity is observed.

Third, not every nanoblock exhibits NDR; for example, the *w* = 120 nm one shows linear response throughout our excitation intensity range (Fig. E4). Below we shall unravel the mechanism of this huge and anomalous nonlinearity.

Silicon is known to exhibit various optical nonlinearities, including parametric processes such as frequency mixing and optical Kerr effect, as well as non-parametric processes such as multiphoton absorption, inelastic scattering (e.g. Raman), free carrier absorption (FCA), photothermal effect (PT), etc. The nonlinearity magnitudes of the first four are all on the order of $10^{-8}$ - $10^{-9}$ μm$^2$/mW,[24,25] and the last two are known to be the most effective to modify silicon's index. Although FCA in silicon is capable to produce index difference as large as 0.1,[26] it requires strong pulsed excitation. Under our continuous-wave excitation, the free carrier density is estimated to be 4 x 10$^{13}$ cm$^{-3}$, and the corresponding $n_2$ value is only $10^{-7}$ μm$^2$/mW,[24] much smaller than the value we observed experimentally. Therefore, photothermal should be the dominating mechanism. Recently, Mie-resonance-enhanced photothermal effect is reported to provide large third-order nonlinearity in sub-100-nm metallic nanostructures.[23,27]

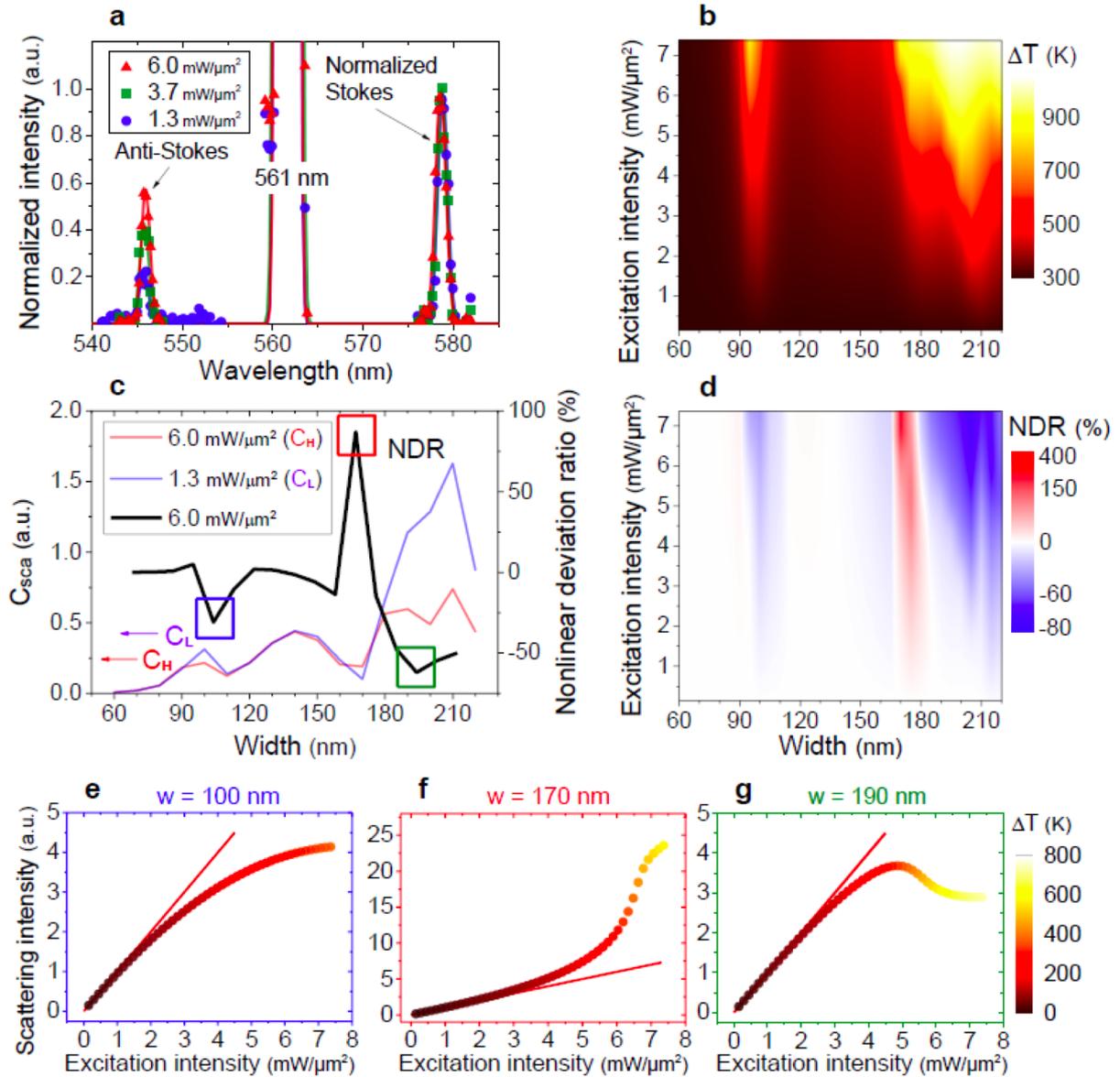

**Fig. 3 | Photothermal-based nonlinearity.**

**a.** The temperature of an individual nanoblock (w = 190 nm) can be in-situ measured by Raman scattering. With increasing excitation intensity, the ratio of anti-Stokes versus Stokes intensity gradually increase, indicating temperature increase up to a few hundred Kelvin (see Methods). **b.** Simulated temperature map of Si nanoblocks versus excitation intensity and width, indicating non-uniform heating up to ~1000K. **c.** Simulated 561-nm scattering cross sections at two excitation intensities (light blue $C_L$: 1.3 mW/μm$^2$; light red $C_H$: 6.0 mW/μm$^2$), and the corresponding NDR (black line, derived from $C_H$ over $C_L$). The size-dependent NDR agrees very well with experimental results in Fig. 2d. **d.** The evolution of NDR versus excitation intensity and nanoblock size. **e, f, g.** Laser intensity dependent scattering for w = 100 nm, 170 nm, and 190 nm nanoblocks, respectively, again agreeing very well with Fig. 2. The color of each dot indicates the equilibrium temperature under photothermal heating.

It is found recently that silicon nanostructures can be very effective optical heaters,[28] especially when excitation is resonant with MQ. From Fig. 2 and Fig. E3, nanoblocks with excitation at MQ resonance (the 190-nm-width one, see Fig. E3b), or excitation at a spectral valley next to MQ resonance (the 170-nm-width one, Fig. E3a), demonstrate better nonlinearity. It is interesting to note that the Q-factor of MQ resonance is only around 20-30, but the photothermal nonlinearity of silicon nanoblock is five order larger than that of bulk. The reason is that Mie resonance enhances nano-silicon absorption by Q = 20-30 times over bulk, but temperature rise can be much more than Q times because the nano-silicon is surrounded by low-thermal-conductance materials (air and quartz). That is, the greatly enhanced photothermal nonlinearity is attributed to not only enhanced absorption from Mie resonance, but also high-efficiency heating due to thermally isolated environment. Note that our reported nonlinear response in an isolated low-Q silicon resonator is much higher than that in a thermally coupled nanostructures on a metasurface with Fano resonance (Q-factor ~1000).[19] The latter provides maximally 30% modulation of transmission at excitation intensity of $10^4$ mW/μm$^2$, while we achieve 400% modulation at a few mW/μm$^2$.

The extraordinary temperature rise is verified through Raman scattering measurement, Fig. 3a unravels a few hundred Kelvin temperature elevation from a single silicon nanostructure (see Methods for derivation), further confirming the existence of photothermal effect. We have characterized the temperature dependent complex refractive index of silicon by ellipsometry (see Fig. E9). In the following, detailed simulation based on photothermal response and scattering of a Mie-resonant silicon nanoblock is carried out, showing outstanding agreement with experiments.

The size- and temperature-dependent absorption cross sections are given in Fig. E10, presenting the need of iterative calculation to derive correct temperature elevation under photothermal effect (see Methods). The iterative result is given in Fig. 3b, where a few hundred Kelvin temperature increase is found, not only showing that silicon nanostructures are indeed efficient heaters, but agreeing well with our Raman experiment. The absorption-induced temperature increase in turn affects refractive index as well as scattering cross section, known as thermo-optic effect, thus leading to nonlinear optical behaviors.

Fig. 3c depicts the 561-nm scattering cross sections at low-intensity (1.3 mW/μm$^2$, light blue curve) and high-intensity (6.0 mW/μm$^2$, light red curve) excitations, manifesting dramatic variation. Their ratio represents simulated size-dependent NDR, i.e. black curve in Fig. 3c, which agrees well with experimental results in Fig. 2d. It is understandable now why the 110-nm and 140-nm nanoblocks exhibit diminishing nonlinearity, since their heating is not significant. Full intensity-dependent evolution of NDR is given in Fig. 3d, showing large NDR indeed corresponds to large temperature elevation.

Further verification with experimental results is provided in Figs. 3e-3g, which are nonlinear scattering of 100-, 170-, and 190-nm nanoblocks versus excitation intensity. Striking similarities to Figs. 2a-2c are found, justifying the correctness of both experiments and the simulations. Therefore, we conclude that Mie-resonance-enhanced photothermal effect[27] is the dominating mechanism of the unexpectedly large nonlinear response in a single silicon nanostructure. It would be an interesting theoretical challenge to find out an analytical expression of photothermal nonlinearity versus Q factor.

One feature of photothermal effect is the sensitivity to the surrounding. Fig. E11 shows the nonlinear response with silicon nanoblocks immersed in glass-index-matching oil, whose thermal conductivity is one order larger than air. Very steep nonlinear variation is found in the intensity dependency, potentially enabling high-contrast all-optical control with a small power variation. More studies would be necessary to investigate the best shape/size/environment for heating nanoparticle, and for inducing maximal photothermal nonlinearity.

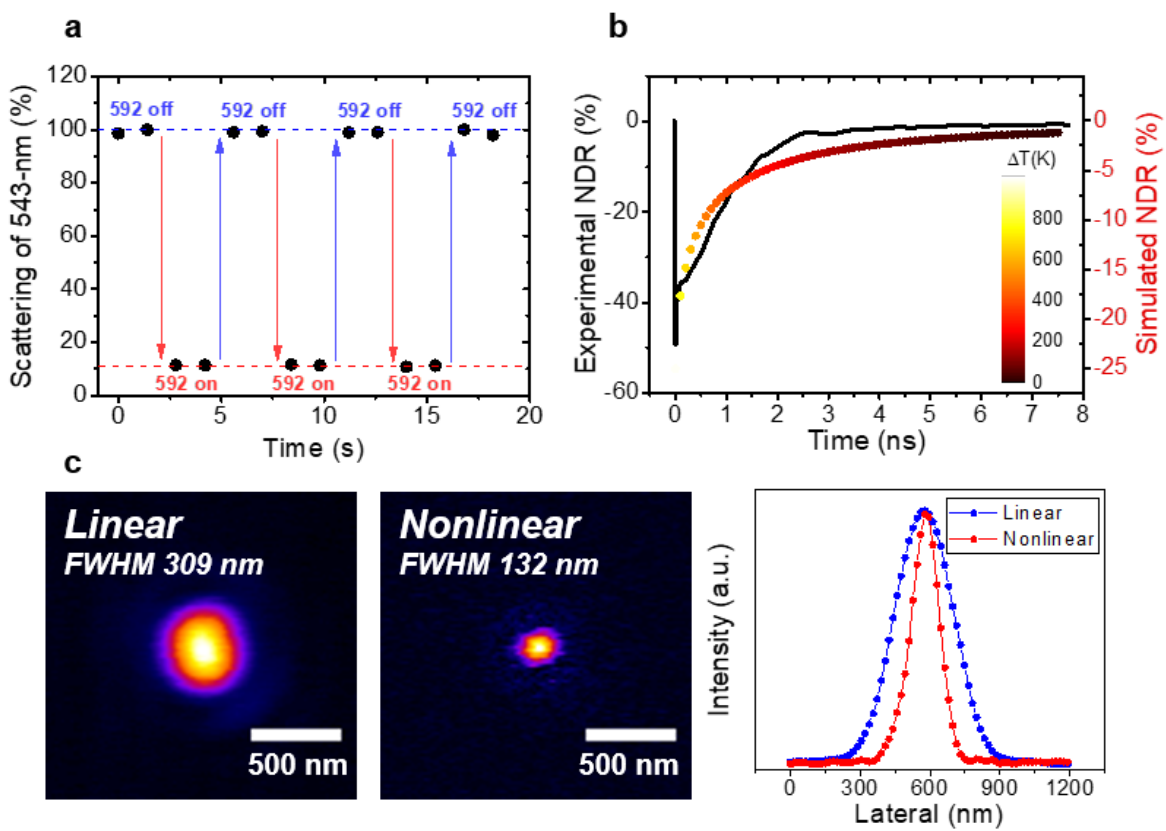

**Fig. 4 | Applications of photothermal nonlinear scattering.**
**a.** All-optical switch on a single nanoblock with simultaneous illumination by a pump (592-nm) and a probe (543-nm) beams. When pump is on, the probe scattering is efficiently turned off. **b.** Temporal evolution of scattering and temperature relaxation in a single nanoblock. Black line is transient pump-probe experimental result, agreeing well with simulation (colored line; colors represent temperature). Both show a relaxation lifetime of ~1 ns. **c.** Resolution enhancement of a silicon nanostructure via nonlinearity. The size of PSF is reduced by more than 2-fold with nonlinear response. Corresponding cross-sectional line profiles are provided in the right panel.

The giant nonlinearity of a single silicon nanostructure that we report here can be applied to various photonic applications, such as ultrasmall all-optical switch and super-resolution imaging on silicon, which is shown in Fig. 4. Fig. 4a demonstrates the all-optical switch, where scattering of a probe beam (at 543 nm) from a single silicon nanostructure (240 x 240 nm) can be efficiently switched off via overlapped excitation of a pump beam (at 592 nm). The modulation depth reaches 90%, and it is fully reversible/repeatable.

One important factor in all-optical switch is speed. Fig. 4b shows the transient response of nonlinear scattering deviation via pump-probe technique (black line, see supplementary methods), and simulation (colored line). Temperature variation of the nanoblock during relaxation is given by the colors. Here we show for the first time, theoretically and experimentally together, that photothermal relaxation of an isolated silicon nanostructure reaches nanosecond, i.e. GHz operation potential, with very large modulation depth that none of previous literature reported. We envision to integrate the reversible photothermal nonlinearity into the field of meta optics, to achieve highly desirable all-optical tunability with a thermally isolated silicon nanostructure. Note that for thermally coupled nanostructures on a metasurface, to avoid heat accumulation, their repetition rate has to be reduced down to kHz.[19]

In the experimental curve, there are apparently two relaxation processes. First, ΔS/S reaches -50% within the pulse duration (~ 1 ps) and subsequently relaxes to -35% within 0.1 ns. This step is dominated by high-density free carriers ($10^{20}$ - $10^{21}$ $cm^{-3}$), whose energy transfer to lattice temperature through Shockley-Read-Hall process and Auger recombination. Next, a second relaxation takes a few nanoseconds to zero NDR (ΔS/S=0). This is primarily attributed to thermal dissipation from Si nanoblock to the surrounding medium. Considering that similar elevated temperature is reached via ultrafast and CW (or long pulse) excitations, the thermal relaxation process after laser off should be equivalent in both cases. More studies are required to understand the slight difference between simulation and experiment in the slow process. Furthermore, optimized responses are expected if more factors such as particle geometries, immersion materials, or phase change effects, are investigated.

There are several possibilities to increase the modulation speed. One is wavelength division multiplexing, which allows simultaneous signal processing at different wavelengths (Fig. 4a) to enhance bandwidth. On the other hand, it is known that at scale less than 100 nm, thermal conductance would be dictated by ballistic condition, i.e. thermal conductance per unit area becomes a constant.[30] Therefore, thermal relaxation time would be directly proportional to its height. In our current experiment, 150-nm height is adopted. Thus, by reducing the height to 10-nm, one order enhancement of speed can be expected. Moreover, the speed may be further enhanced by replacing the substrate with a material of higher thermal conductivity. These will be our future directions.

Fig. 4c shows a potential application of all-optical switch, i.e. super-resolution imaging. It is well known that the capability to precisely control light emission on/off leads to significant resolution enhancement,[29] as we have previously shown for plasmonic nanostructures.[27] On the other hand, nonlinear response itself also enables significant resolution enhancement (see supplementary methods).[31] Here we demonstrate 2.3x resolution enhancement in Fig. 4c, i.e. beyond diffraction limit, with a 100 nm nanoblock. The nanostructure exhibits strong nonlinear response (saturation of scattering in Fig. 2a). With a Gaussian focus, nonlinear response should start from the center of PSF, and thus by extracting the nonlinear part, the resulting PSF becomes smaller than its linear counterpart. It is to our knowledge, the first demonstration of super-resolution microscopy on silicon materials, and can be applied not only to label-free silicon nanostructure observation, but also further into biomedical applications with silicon nanoparticles.

In summary, we discovered large and fast photothermal nonlinearity of Si nanostructures, enabled by Mie-resonance enhanced absorption and thermally isolated efficient heating. The nonlinear coefficient ($n_2$) is five orders larger over bulk silicon photothermal nonlinearity, and is much larger compared to all previous nonlinear silicon reports, thus allowing more than +400% ~ -70% nonlinear deviation of scattering under CW illumination. Transient measurements and simulations revealed nanosecond thermal dissipation time, without sacrificing the large NDR. Our results open up a new direction in nonlinear silicon nanophotonics that can be applied toward high-speed, high-contrast all-optical switch in nanoscale, as well as super-resolution imaging of silicon.

# References

1. Soref, R. The Past, Present, and Future of Silicon Photonics. *IEEE J. Sel. Top. Quantum Electron.* **12**, 1678–1687 (2006).

2. Reed, G. T., Mashanovich, G., Gardes, F. Y. & Thomson, D. J. Silicon optical modulators. *Nat. Photonics* **4**, 518–526 (2010).

3. Lin, D., Fan, P., Hasman, E. & Brongersma, M. L. Dielectric gradient metasurface optical elements. *Science* **345**, 298–302 (2014).

4. Kivshar, Y. All-dielectric meta-optics and non-linear nanophotonics. *National Science Review* **5**, 144–158 (2018).

5. Leuthold, J., Koos, C. & Freude, W. Nonlinear silicon photonics. *Nat. Photonics* **4**, 535–544 (2010).

6.  Li, Q., Davanço, M. & Srinivasan, K. Efficient and low-noise single-photon-level frequency conversion interfaces using silicon nanophotonics. *Nat. Photonics* **10**, 406–414 (2016).

7.  Staude, I. & Schilling, J. Metamaterial-inspired silicon nanophotonics. *Nat. Photonics* **11**, 274–284 (2017).

8.  Horvath, C., Bachman, D., Indoe, R. & Van, V. Photothermal nonlinearity and optical bistability in a graphene–silicon waveguide resonator. *Opt. Lett.* **38**, 5036 (2013).

9.  Atabaki, A. H. *et al.* Integrating photonics with silicon nanoelectronics for the next generation of systems on a chip. *Nature* **556**, 349–354 (2018).

10. Kuznetsov, A. I., Miroshnichenko, A. E., Brongersma, M. L., Kivshar, Y. S. & Luk'yanchuk, B. Optically resonant dielectric nanostructures. *Science* **354**, (2016).

11. Kuznetsov, A. I., Miroshnichenko, A. E., Fu, Y. H., Zhang, J. & Luk'yanchuk, B. Magnetic light. *Sci. Rep.* **2**, 492 (2012).

12. Fu, Y. H., Kuznetsov, A. I., Miroshnichenko, A. E., Yu, Y. F. & Luk'yanchuk, B. Directional visible light scattering by silicon nanoparticles. *Nat. Commun.* **4**, 1527 (2013).

13. Moitra, P. *et al.* Large-Scale All-Dielectric Metamaterial Perfect Reflectors. *ACS Photonics* **2**, 692–698 (2015).

14. Wang, L. *et al.* Grayscale transparent metasurface holograms. *Optica* **3**, 1504 (2016).

15. Hafezi, M., Mittal, S., Fan, J., Migdall, A. & Taylor, J. M. Imaging topological edge states in silicon photonics. *Nat. Photonics* **7**, 1001–1005 (2013).

16. Nagasaki, Y., Suzuki, M. & Takahara, J. All-Dielectric Dual-Color Pixel with

Subwavelength Resolution. *Nano Lett.* **17**, 7500–7506 (2017).

17. Li, M., Zhang, L., Tong, L.-M. & Dai, D.-X. Hybrid silicon nonlinear photonics [Invited]. *Photonics Research* **6**, B13 (2018).

18. Shalaev, M. I. *et al.* High-Efficiency All-Dielectric Metasurfaces for Ultracompact Beam Manipulation in Transmission Mode. *Nano Lett.* **15**, 6261–6266 (2015).

19. Yang, Y. *et al.* Nonlinear Fano-Resonant Dielectric Metasurfaces. *Nano Lett.* **15**, 7388–7393 (2015).

20. Shcherbakov, M. R. *et al.* Enhanced third-harmonic generation in silicon nanoparticles driven by magnetic response. *Nano Lett.* **14**, 6488–6492 (2014).

21. Shcherbakov, M. R. *et al.* Ultrafast All-Optical Switching with Magnetic Resonances in Nonlinear Dielectric Nanostructures. *Nano Lett.* **15**, 6985–6990 (2015).

22. Nagasaki, Y., Suzuki, M., Hotta, I. & Takahara, J. Control of Si-Based All-Dielectric Printing Color through Oxidation. *ACS Photonics* **5**, 1460–1466 (2018).

23. Chen, Y.-T. *et al.* Study of Nonlinear Plasmonic Scattering in Metallic Nanoparticles. *ACS Photonics* **3**, 1432–1439 (2016).

24. Borghi, M., Castellan, C., Signorini, S., Trenti, A. & Pavesi, L. Nonlinear silicon photonics. *J. Opt.* **19**, 093002 (2017).

25. Lin, Q., Painter, O. J. & Agrawal, G. P. Nonlinear optical phenomena in silicon waveguides: modeling and applications. *Opt. Express* **15**, 16604–16644 (2007).

26. Shcherbakov, M. R. *et al.* Ultrafast all-optical tuning of direct-gap semiconductor metasurfaces. *Nat. Commun.* **8**, 17 (2017).

27. Wu, H.-Y. *et al.* Ultrasmall all-optical plasmonic switch and its application to superresolution imaging. *Sci. Rep.* **6**, 24293 (2016).

28. Zograf, G. P. *et al.* Resonant Nonplasmonic Nanoparticles for Efficient Temperature-Feedback Optical Heating. *Nano Lett.* **17**, 2945–2952 (2017).

29. Hell, S. W. Far-field optical nanoscopy. *Science* **316**, 1153–1158 (2007).

30. Anufriev, R., Gluchko, S., Volz, S. & Nomura, M. Quasi-Ballistic Heat Conduction due to Lévy Phonon Flights in Silicon Nanowires. *ACS Nano* **12**, 11928–11935 (2018).

31. Chu, S.-W. *et al.* Measurement of a saturated emission of optical radiation from gold nanoparticles: application to an ultrahigh resolution microscope. *Phys. Rev. Lett.* **112**, 017402 (2014).

**Acknowledgements**

This work was supported by the Outstanding Young Scholarship Project of Ministry of Science and Technology, Taiwan, under grant MOST-105-2628-M-002-010-MY4 (SWC), as well as other projects MOST-107-2321-B-002-009 (SWC), and 108-2112-M-001-027 (KHL). SWC acknowledges the generous award from the Foundation for the Advancement Outstanding Scholarship. This work was also supported by the Photonics Center at Osaka University and Japan Society for the Promotion of Science (JSPS) Core-to-Core Program, A. Advanced Research Networks. YN was supported by a JSPS Research Fellowship for Young Scientists. A part of this work was supported by the "Nanotechnology Platform Project (Nanotechnology Open Facilities in Osaka University)" of the Ministry of Education, Culture, Sports, Science and Technology, Japan [No. F-17-OS-0011 and S-17-OS-0011].

We would like to thank Dr Yu-Ming Chang for his helpful discussion on Raman scattering, and Mr. Guan-Jie Huang, Mr. Yu-Feng Chien for their helpful assistance in experiments. The ellipsometry measurement was kindly supported by Prof. Hsiang-Lin Liu and Mr. Hsiao-Wen Chen. We also would like to thank Shin-Etsu Chemical Co., Ltd. for donating high-quality silicon-on-quartz substrates.